\documentclass[conference]{IEEEtran}
\IEEEoverridecommandlockouts
\usepackage[switch]{lineno}
\usepackage{cite}
\usepackage{amsmath,amssymb,amsfonts}
\usepackage{algorithmic}
\usepackage{graphicx}
\usepackage{textcomp}
\usepackage{xcolor}
\def\BibTeX{{\rm B\kern-.05em{\sc i\kern-.025em b}\kern-.08em
    T\kern-.1667em\lower.7ex\hbox{E}\kern-.125emX}}

\usepackage{booktabs}
\usepackage{multirow}
\usepackage{caption}
\usepackage{subcaption}
\usepackage{amsthm}
\usepackage{bbm}
\usepackage{dcolumn}
\usepackage{import}
\usepackage{pdflscape}
\usepackage{float}%
\usepackage{afterpage}
\usepackage{enumitem}
\newtheorem{definition}{Metric}

\usepackage{tabularx,ragged2e}

\begin{document}

\title{%
Leveraging Complementary AI Explanations \\to Mitigate Misunderstanding in XAI%
}
\makeatletter
\newcommand{\linebreakand}{%
  \end{@IEEEauthorhalign}
  \hfill\mbox{}\par
  \mbox{}\hfill\begin{@IEEEauthorhalign}
}
\makeatother

\author{
\IEEEauthorblockN{Yueqing Xuan}
\IEEEauthorblockA{\textit{School of Computing Technologies} \\
\textit{RMIT University, Australia}\\
yueqing.xuan@student.rmit.edu.au}
\and
\IEEEauthorblockN{Kacper Sokol}
\IEEEauthorblockA{\textit{Department of Computer Science} \\
\textit{ETH Zurich, Switzerland}\\
kacper.sokol@inf.ethz.ch}
\and
\IEEEauthorblockN{Mark Sanderson, Jeffrey Chan}
\IEEEauthorblockA{\textit{School of Computing Technologies} \\
\textit{RMIT University, Australia}\\
\{mark.sanderson, jeffrey.chan\}@rmit.edu.au}
}

\maketitle
\begin{abstract}%
Artificial intelligence explanations %
can make %
complex predictive models more comprehensible. %
To be effective, however, they should %
anticipate and mitigate possible misinterpretations, e.g., arising %
when users infer incorrect information that is not explicitly conveyed.  %
To this end, %
we propose \emph{complementary explanations} -- a novel method that pairs explanations to compensate for %
their respective %
limitations. %
A complementary explanation %
adds %
insights that clarify potential misconceptions stemming from the primary explanation while ensuring their coherency and avoiding redundancy. %
We introduce a framework for designing and evaluating complementary explanation pairs based on %
pertinent qualitative properties and quantitative metrics. %
Our approach allows to %
construct complementary explanations that minimise %
the chance of their misinterpretation.%
\end{abstract}%
\begin{IEEEkeywords}%
machine learning, artificial intelligence, explainability, intelligibility, comprehension, evaluation, human-centred.%
\end{IEEEkeywords}%

\section{Introduction}
Artificial intelligence (AI) explanations assist users in interpreting the general functioning as well as selected details of complex predictive models. %
When those AI models are opaque, their explanations become a crucial bridge ensuring that users can better understand them. %
\emph{Effective} explanations should not only maximise comprehension of insights that they convey but also mitigate their misinterpretation. %
While relevant research has mainly focused on the former task -- by improving the intelligibility and informativeness of AI explanations -- a critical, yet often overlooked, challenge remains in minimising user misunderstanding, in particular when explainees infer spurious information that an explanation does not provide. %

The key cognitive bias contributing to this phenomenon is %
\emph{the illusion of explanatory depth}, leading users to believe that they understand a system in greater detail than they actually do~\cite{chromik2021think,byrne2023good}.  %
This can result in flawed judgment, automation misuse and miscalibrated trust, where users either over-rely on or unjustifiably dismiss AI output~\cite{parasuraman1997humans,jacovi2021formalizing, ahn2024impact}.  %
Prior work has shown that explanations for which users struggle to identify unspecified information -- thus realise their limitations -- are particularly prone to misinterpretation~\cite{xuan2025comprehension}. %
To ensure responsible AI adoption, it is critical not only to enhance user comprehension of what an explanation conveys but also to prevent users from making unwarranted generalisations about information that remains unknown.

A common approach to mitigate misinterpretation is to explicitly inform users about the limitations of an explanation and caution them against drawing conclusions from missing information~\cite{van2021effect}. 
However, given the breadth of missing information in an explanation, listing all such omissions is neither practical nor effective as it risks overwhelming users with excessive detail. %
Another approach is to make explanations interactive, %
which allows users to independently explore AI models~\cite{sokol2018glass}. %
While this strategy provides access to richer information, it does not guarantee that users will develop the necessary understanding; also, we still lack systematic methodology for constructing interactive explanations while preventing their individual misinterpretation. %

In this paper we introduce \emph{complementary explanations} to mitigate user misunderstanding of AI explanations. 
A complementary explanation provides additional information specifically designed to address common misinterpretations and clarify missing details that users are likely to incorrectly infer from the primary explanation. 
Intelligible explanations, while effective in conveying explicated information, can paradoxically be misleading if they unintentionally cause users to overgeneralise the information that they provide~\cite{xuan2025comprehension}.  %
To this end, we propose a framework for systematically identifying misleading aspects of an explanation and pairing it with a complementary explanation that addresses these gaps. %

Our framework %
is applied in three steps. %
First, we identify the explicitly conveyed and unspecified information of an AI explanation. %
Second, we select its complementary explanation based on four desiderata: %
(1)~degree of \emph{novelty}; %
(2)~difference in \emph{granularity}; %
(3)~extent of \emph{non-redundancy}, i.e., capacity to provide additional information rather than reiterate existing insights; and %
(4)~level of \emph{coherency}, i.e., ability to deliver sufficient shared context to help users integrate multiple explanations into a meaningful bigger picture. %
Third, we %
quantify the degree of complementarity between explanations using dedicated metrics. %
This structured approach enables %
constructing principled %
complementary explanation pairs %
that %
enhance user comprehension while minimising %
misinterpretation risk. %

\section{Background and Related Work}%

When users interact with AI models through their explanations it is crucial that they understand how to correctly interpret such insights~\cite{sokol2024does}. %
A common approach is to explicitly communicate the limitations of AI explanations, %
e.g.,
indicate the information that remains unspecified by explanatory artefacts~\cite{van2021effect}. %
Consider
an AI model assisting doctors in predicting diabetes risk; a local explanation tailored to a particular patient might state the limitation of its information in a form of a disclaimer: ``the explanation applies exclusively to this patient and does not imply that \emph{glucose} is the most important factor across all the cases''. %
Explicitly communicating explanation scope helps to prevent its overgeneralisation. %

Simply disclosing the limitations of an explanation may nonetheless be insufficient to curb its misinterpretation. 
This is because explanations tend to vary in scope (e.g., global, local or sub-space) and information content (e.g., feature influence or counterfactual insight)~\cite{guidotti2018survey}. %
It is thus impractical to identify and communicate all the unspecified aspects of explanatory insights. %
Moreover, this approach will only succeed if users can understand and operationalise such details. %
Additionally, while the limitation disclosure narrows down the scope of consideration and reduces the required cognitive effort, it does not facilitate richer understanding of AI models and can potentially erode user trust and engagement. %

Another strategy is to allow users to interact with explanations (without supervision) -- e.g., through a dynamic interface~\cite{cheng2019explaining} -- helping them to explore an AI model and incrementally expand their knowledge. %
This approach, however, %
cannot guarantee that users will learn what is necessary to achieve the desired level of comprehension. %
Furthermore, the sequence in which explanations are presented affects how users process information and could lead to inconsistent interpretations~\cite{kaur2022sensible}. %
Thus providing interactive explanatory information without any underlying structure risks confusing or overwhelming users, possibly leading to unexpected misinterpretations. %

Complementary explanations address such challenges by combining structured (interactive) exploration with carefully selected explanatory information. %
They present different yet coherent insights that guide users through the explainability process while minimising cognitive load. %
Our complementary explanations therefore align with the concept of \emph{explanations as a social practice} by facilitating interaction and co-construction~\cite{rohlfing2020explanation}. %
While current research has explored unifying different explanation types -- e.g., combining dataset analysis with global feature importance~\cite{bhattacharya2024exmos} and contextualising local feature attribution with partial dependence plots~\cite{bove2022contextualization} -- %
it is mostly limited to preselected pairs of explanations in isolated contexts, lacking systematic guidelines on selecting and integrating various explanations. %
Our work addresses this gap by demonstrating how to design complementary explanations that minimise misunderstanding and support learning. %

\begin{table}[t]
    \centering
    \footnotesize
    \caption{Selected explanation types -- split into three groups: global, local and sub-space -- and their information scope. %
    These explanations can be generated with well-established, model-agnostic tools such as LIME, SHAP and PDP. %
    }%
    \setlength{\tabcolsep}{2.605pt}
    \begin{tabular}{@{}>{\RaggedRight}p{2.045cm}>{\RaggedRight}p{3.2315cm}>{\RaggedRight}p{3.2315cm}@{}}%
    \toprule
       \textbf{Explanation} & \textbf{Explicated Information} & \textbf{Unspecified Information} \\ \midrule
       partial dependence~plot & overall effect of a feature on the model's output & feature interaction \& instance-level insights \\ \cmidrule(lr){2-3}%
       (surrogate)~decision~rules~\&~trees & rules \& trees approximating overall model behaviour %
       & surrogate fidelity \& %
       feature relationship \\ \cmidrule(lr){2-3}%
       feature importance & overall importance of features for predictions & instance-wise importance \& feature interaction \\ \cmidrule(lr){2-3} %
       data distribution analysis & dataset statistics, outliers \& feature distribution & context of how feature distribution affects~predictions \\ \midrule
        counterfactual \& decision surface & changes required to alter a specific prediction & how changes affect other predictions \\ \cmidrule(lr){2-3}
        feature attribution & contribution of individual features to a prediction & feature interaction \& global feature importance \\ \cmidrule(lr){2-3}
        nearest neighbours & closest training points to a given input & global patterns \& model-level insights \\ \cmidrule(lr){2-3}
        influence function & influence of a data point on a prediction & how data \& features interactions impact~predictions\\ \midrule%
        prototypes \& criticisms & representative examples \& edge cases or exceptions & broader regional patterns \& model generalisability \\ \cmidrule(lr){2-3}%
        regional feature importance & feature importance within a specific data sub-space & global feature importance \& variation across spaces\\
    \bottomrule
    \end{tabular}
    \label{tab:summary_exp}
\end{table}

\section{Complementary Explanation Framework}

Our complementary explanation framework builds upon existing explanation types, %
aligning them in a systematic way. %

\subsection{Identifying Information}
 
Complementary explanations match different explanation types by identifying their pairs that offer complementary information to compensate for their respective shortcomings and limitations. %
When linked to a primary explanation, the complementary explanation adds information that the former does not fully convey and is likely to be misinterpreted. %
For example, user studies have shown that lay people often infer local feature attribution from decision surface visualisation and counterfactuals~\cite{xuan2025comprehension}. By presenting feature importance alongside these explanation types we can prevent such misconceptions. In this context, local feature attribution is complementary to decision surface visualisation and counterfactuals since it minimises the chances of users inferring incorrect insights that the latter do not specify. %

Given the variety of available explanation types, we require a systemic guideline to assess their information scope -- whether explicated or unspecified -- to apply our framework. %
Specifically, we need to know: %
    (1)~what information an explanation \emph{communicates}, and whether the explicated information is intelligible; and %
    (2)~what information an explanation \emph{misses}, and what \emph{unspecified} insight is the most likely to be misconstrued by users. %
Table~\ref{tab:summary_exp} provides answers to these questions for a selection of representative AI explanation types. %

\subsection{Design Principles}\label{sec:design-principle}

Identifying the information scope of different explanations allows us to %
determine their complementarity, %
which we chart across four dimensions: %
novelty, granularity, non-redundancy and coherency.

\paragraph{Novelty}%
    The complementary explanation should provide information beyond what is communicated by the corresponding primary explanation. %
    For example, local feature attribution can be complemented by data distribution analysis, which contextualises feature values of an individual within the broader population. %
    Since the former explanation type lacks insights about (a)typicality of feature values, the latter addresses this shortcoming by showing how these attribute values align with or deviate from the modelled population. %
    This %
    extension %
    connects local insights to broader data patterns. %
    
\paragraph{Granularity}
    The complementary explanation can present a different level of granularity, providing either a broader or a more detailed view. For example, a high-level explanation of model behaviour may highlight that \emph{location} is the most important feature in predicting house pricing, while a local explanation might reveal that \emph{size} is more influential for a specific home. %
    Providing explanations at multiple granularity levels helps users to engage with AI models on various analytical planes. %
    This combination also caters to different cognitive approaches: inductive reasoning, which generalises from specific instances to broader patterns, and deductive reasoning, which applies general rules to specific outcomes~\cite{wang2019designing}. %
\paragraph{Non-redundancy}
    The complementary explanation should avoid excessive repetition of information communicated by the primary explanation. %
    For example, decision rules show conditions leading to specific outcomes, %
    and
    counterfactuals describe how to tweak input features to change outcomes. %
    Since both focus on feature values that determine the output of a model, they are largely redundant in conjunction. %
    
\paragraph{Coherency}
    An explanation pair should %
    share context to %
    help users integrate these insights into a coherent mental representation of an AI model. %
    Coherency ensures that explanations are perceived as interconnected parts of a whole rather than isolated pieces of information. An example of incoherency would be pairing a counterfactual with global feature importance. %
    While the former focuses on instance-level changes, %
    the latter ranks features by their overall impact. %
    The lack of a direct connection between the individual and global perspectives can prevent users from reconciling these explanations to form a unified understanding of an AI model. %

\begin{figure}[t]
    \centering
    \begin{subfigure}[b]{0.325\linewidth}
        \centering
        \includegraphics[clip,width=\textwidth]{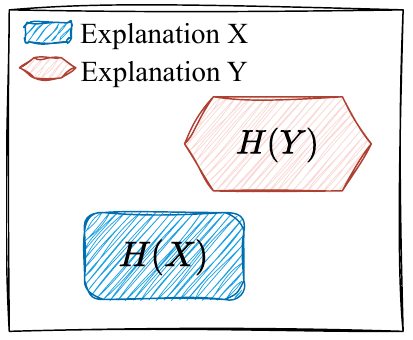}
        \caption{None.}\label{fig:no-mutual-info}%
    \end{subfigure}
    \hfill
    \begin{subfigure}[b]{0.325\linewidth}
        \centering
        \includegraphics[clip,width=\textwidth]{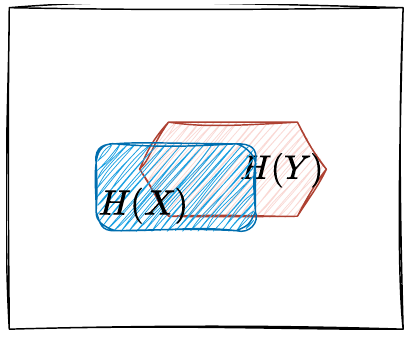}
        \caption{Excessive.}\label{fig:too-much-mutual} %
    \end{subfigure}
    \hfill
    \begin{subfigure}[b]{0.325\linewidth}
        \centering
        \includegraphics[clip,width=\textwidth]{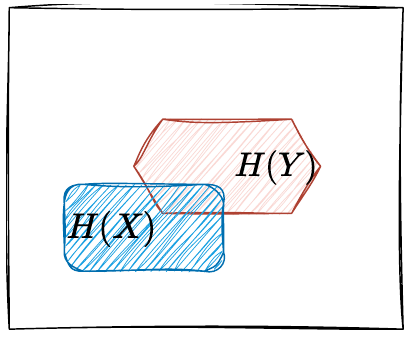}
        \caption{Optimal.}\label{fig:good-mutual}%
    \end{subfigure}
    \caption{%
    \emph{Mutual Information} examples for explanations $X$~\&~$Y$.%
    }
    \label{fig:mutual-info}
\end{figure}

\subsection{Evaluation Metrics}

Next, we introduce three metrics to %
quantify complementarity of explanation types and guide their selection. %

\begin{definition}[Information Richness]\label{def:info-richness}
    Information richness $H(X)$ of explanation $X$ %
    measures the amount of intelligible information that $X$ conveys about the behaviour of an AI model. %
\end{definition}

The value of \emph{Information Richness} -- Metric~\ref{def:info-richness} -- depends on factors such as the number of features included in $X$, the modality and size of $X$, and its presentation structure. %
For example, %
the length of an explanation or the number of new concepts that it introduces as well as the overall size of the feature set can be used to this end~\cite{narayanan2018humans,lage2019human,xuan2025comprehension}. %
Similarly, the cognitive load of visual AI explanations can be quantified by analysing feature counts and visual trends~\cite{abdul2020cogam}. %
Since overly complex explanations can cause cognitive overload~\cite{bo2024incremental}, $H(X)$ exhibits sub-linear growth as explanation content increases, reflecting the diminishing returns when additional information becomes harder to process. %

\begin{definition}[Mutual Information]\label{def:mutual-info}
Mutual information $I(X; Y) = H(X)\cap H(Y)$ quantifies the amount of information shared between explanations $X$ and $Y$. %
\end{definition}

\emph{Information Richness} characterises a single explanation. When aligning multiple explanations, we use \emph{Mutual Information} -- Metric~\ref{def:mutual-info} -- to measure the degree of redundancy between them.
As discussed in Sect.~\ref{sec:design-principle}, effective complementary explanations should have small content overlap (non-redundancy) while preserving some shared information to maintain thematic coherency. Figure~\ref{fig:mutual-info} illustrates this concept. %
Ideally, $I(X; Y)$ should be low but not zero; when $I(X; Y) \approx 0$ users may struggle to reconcile the explanations and develop cohesive understanding (Fig.~\ref{fig:no-mutual-info}).  %
As an example, consider counterfactuals paired with global feature importance. %

When $I(X; Y)$ is excessive (Fig.~\ref{fig:too-much-mutual}), such as decision rules and counterfactuals, the second explanation adds little new insight but processing it requires extra cognitive effort. %
The optimal amount of \emph{Mutual Information} (Fig.~\ref{fig:good-mutual}) delivers novel insights and facilitates coherent comprehension. %
For example, %
pairing counterfactuals with local feature attribution offers both diagnostic (what contributed to the decision) and actionable (what to change to alter the outcome) insights. %

\begin{figure}[t]
    \centering
    \begin{subfigure}[b]{0.325\linewidth}
        \centering
        \includegraphics[clip,width=\textwidth]{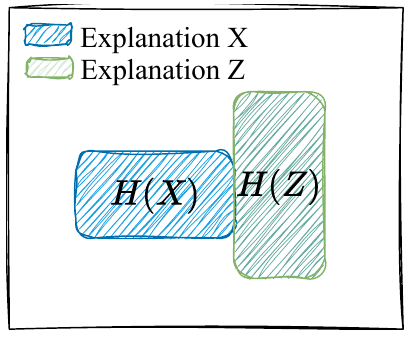}
        \caption{Maximum.}\label{fig:max-info-gain}%
    \end{subfigure}
    \hfill
    \begin{subfigure}[b]{.325\linewidth}
        \centering
        \includegraphics[clip,width=\textwidth]{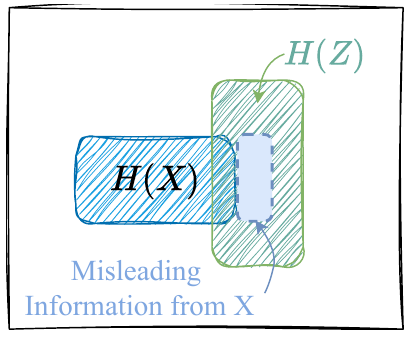}
        \caption{Desirable.}\label{fig:good-info-gain}%
    \end{subfigure}
    \hfill
    \begin{subfigure}[b]{.325\linewidth}
        \centering
        \includegraphics[clip,width=\textwidth]{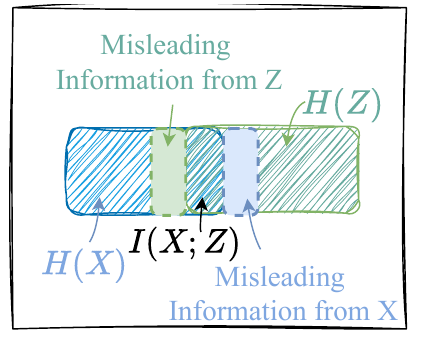}
        \caption{Complementary.}\label{fig:complete-comp}%
    \end{subfigure}
    \caption{%
    \emph{Information Gain} examples for explanations $X$~\&~$Z$.%
    }%
    \label{fig:info-gain}
\end{figure}

\begin{definition}[Information Gain]\label{def:info-gain}
Information gain $IG(Y, X) = H(Y) - I(X;Y)$ quantifies the amount of new information that explanation $Y$ provides %
in the context of explanation $X$. %
\end{definition}

While \emph{Mutual Information} captures how well a complementary explanation aligns with the primary one, we also need a metric to assess its ability to offer new insights and mitigate any potential misunderstanding caused by the primary explanation. %
\emph{Information Gain} -- Metric~\ref{def:info-gain} -- captures insight novelty and granularity difference that the complementary explanation introduces -- see Fig.~\ref{fig:info-gain}. %
Maximum \emph{Information Gain} is achieved when there is no overlap between explanations, i.e., $I(X; Y)=0$, and $H(Y)$ is large enough to ensure sufficient \emph{Information Richness} (Fig.~\ref{fig:max-info-gain}). %
This, however, is suboptimal as %
some shared information is needed for coherency. %
It is more desirable for %
the insights provided by $Y$ to fill the %
informational gap left by $X$ to mitigate any misunderstanding (Fig.~\ref{fig:good-info-gain}). %
In this instance, the complementary explanation not only provides novel insights but also rectifies potential misconceptions. %
Furthermore, maintaining sufficient shared information helps users to merge the explanations into cohesive understanding and reduces cognitive load. %

Based on \emph{Mutual Information} and \emph{Information Gain}, we introduce \emph{complementary} explanations -- illustrated in Fig.~\ref{fig:complete-comp}. %
Each explanation in this pair mitigates potential misconceptions introduced by the other.
Their shared information fosters coherent understanding, ensuring that they work together to broaden comprehension. %
For example, consider local feature attribution and counterfactual explanations in the context of predicting heart attack risk. %
The former %
highlights the contribution of the most influential attributes  -- e.g., cholesterol, blood pressure and exercise frequency -- to a model's output; %
it shows %
why a prediction was made %
but lacks information about %
how changes to feature values affect this outcome. %
A user may thus falsely believe that lowering cholesterol alone is sufficient to reduce their heart attack risk since it is the biggest factor. %
This interpretation, however, overlooks possible feature interactions. %

Counterfactual explanations fill this gap by suggesting the smallest actionable feature change, e.g., reducing blood pressure in combination with increasing exercise frequency, %
which implicitly accounts for attribute interaction. %
Thus rather than applying a big change to the most important feature (cholesterol), small adjustments to multiple attributes bring the desired outcome. %
When provided in isolation, %
counterfactuals %
may nonetheless mislead a user into believing that blood pressure and exercise frequency are the most significant factors, causing them to overlook the contribution of each attribute independently~\cite{xuan2025comprehension}. %

Feature attribution addresses the limitation of counterfactuals and clarifies that cholesterol is still the most influential factor. %
Together, these explanations address each other's limitations. 
Their shared information -- common explainability scope, context and content (i.e., feature set) -- brings them together into a coherent narrative. %
In summary, complementary explanations are mutually enriching, helping to improve overall user understanding of AI models.  %

\section{Conclusion and Future Work}

Ensuring that AI explanations do not mislead users about the information they do not communicate is crucial. This paper introduced a framework and guidelines for designing complementary explanations that minimise the chances of any such misunderstanding. %
It also defined criteria to quantify the complementarity of explanations and identified complementary explanation pairs that address each other's limitations. %

Future work will focus on user studies %
aimed at empirical validation of our framework. %
Specifically, we will investigate if complementary explanations reduce misconceptions more effectively than simply disclosing the limitations of each individual explanation.  %
Accounting for user demographics will be crucial when applying our framework as the perceived complementarity of explanation types may vary across stakeholders~\cite{ehsan2024xai}. %
We further plan to examine the impact of complementary explanations on user trust and cognitive load as overly complex explanation pairs may adversely affect both.

\section*{Acknowledgements}
This research was conducted by the ARC Centre of Excellence for Automated Decision-Making and Society (project number CE200100005), funded by the Australian Government through the Australian Research Council. %
Additional support was provided by the Hasler Foundation (grant number 23082).

\bibliographystyle{IEEEtran}
\bibliography{references}

\end{document}